\documentclass[10pt,twocolumn,superscriptaddress,aps,prd]{revtex4-1}
\usepackage{graphicx}
\usepackage{subfigure}
\usepackage{dcolumn}\usepackage{bm}
\usepackage[usenames]{color}
\usepackage{amsmath}
\usepackage{amssymb}
\usepackage{url} 
\usepackage{hyperref}

\newcommand{\f}{\frac}
\newcommand{\beq}{\begin{equation}}
\newcommand{\eeq}{\end{equation}}
\begin{document}
\topmargin 0.0001cm
\title{Probing Leptophilic Dark Sectors at Electron Beam-Dump Facilities}  

\newcommand*{\INFNGE}{Istituto Nazionale di Fisica Nucleare, Sezione di Genova, 16146 Genova, Italy}
\newcommand*{\UNIGE}{Universit\'a degli Studi di Genova, 16126 Genova, Italy}
\newcommand*{\BUNI}{Physics Department, Boston University, Boston, 02215, USA}

\author {L.~Marsicano} 
\affiliation{\INFNGE}
\affiliation{\UNIGE}
\author {M.~Battaglieri} 
\affiliation{\INFNGE}
\author {A.~Celentano} 
\affiliation{\INFNGE}
\author {R.~De Vita} 
\affiliation{\INFNGE}
\author{Yi-Ming Zhong}
\affiliation{\BUNI}

\date{\today}
\begin{abstract}
Medium-energy electron beam-dump experiments provide an intense sources of secondary muons. These particles can be used to search for muon-coupling light dark scalars that may explain the $(g-2)_\mu$ anomaly. We applied this idea to SLAC E137 experiment deriving new exclusion limits  and evaluated the  expected sensitivity for the planned Jefferson Lab BDX experiment (in case of  null result report). The calculation is based on numerical simulations that include a realistic description of secondary muons generation in the dump, dark scalar  production, propagation, and decay, and, finally,  the decay products (electrons, positrons, or photons) interaction with the detector. 
For both experiments,  exclusion limits were extended to cover a broader area in the scalar-to-muon coupling vs. scalar mass parameter space. This study demonstrates that electron beam-dump experiments have an enhanced sensitivity to new physics in processes that are usually studied using proton beams.

\end{abstract}
\maketitle
\section{\label{sec:intro} Introduction}
Fifteen years since it was firstly proposed~\cite{PhysRevLett.92.161802}, the discrepancy between the measured value of the anomalous magnetic moment of muon and its Standard Model (SM) prediction remains unexplained. The so-called ``$(g-2)_\mu$ anomaly'' has provided a strong motivation for light hidden particles searches, opening a window to new physics beyond the Standard Model (SM). Popular candidates such as dark photons and dark Higgs have been tightly scrutinized as a possible explanation for such anomaly. However, this hypothesis has been excluded by existing measurements (see e.g. ~\cite{Battaglieri:2017aum,Krnjaic:2015mbs}). Nevertheless, other solutions, such as a new light particle that dominantly couples to muons, remain viable and deserve attention. In this paper, we  examine models with a new  leptophilic dark scalar, tuned to explain the $(g-2)_\mu$ anomaly. We employ a simplified model framework, with the effective Lagrangian for the dark scalar field $S$ written as:
\begin{equation}
\mathcal{L} \supset \frac{1}{2} (\partial_\mu S)^2 -\frac{1}{2} m_S^2 S^2 - \sum_{\ell = e, \mu, \tau} g_\ell S \bar{\ell} \ell \; \; \;,
\label{eq:scalarL}
\end{equation}
where $S$ is a real scalar field and the coupling between $S$ and the SM leptons, $g_\ell$, is restricted to be
\begin{equation}
\label{med}
  \left \{
  \begin{tabular}{l}
     (Lepton-specific) \quad $g_e:g_\mu:g_\tau = m_e: m_\mu : m_\tau$\\
      (Muon-specific) \quad $g_\mu \neq 0,\, g_e = g_\tau = 0$.
  \end{tabular}
\right.
\end{equation}
Such effective Lagrangian may originate from an effective gauge-invariant dimension 5 operator~\cite{Chen:2015vqy,Batell:2016ove, Chen:2017awl,Batell:2017kty}:
\begin{equation}
\frac{c_i}{\Lambda} S \bar{L}_i H E_i
\end{equation}
where $H$, $L$, $E$ are the SM Higgs doublet, lepton doublet, and lepton singlet respectively. $\Lambda$ is the new physics scale and $c_i$ are the Wilson coefficients for flavor $i$. Although the relative size of $c_i$ and the effective coupling $g_i$ are {\it a priori} undetermined, one may naturally expect  that their values are regulated by the principle of Minimal Flavor Violation (MFV)~\cite{DAmbrosio:2002vsn} and therefore proportional to the Yukawa couplings $y_i$. This leads to the coupling relations dubbed as ``lepton-specific''. On the other end, one may go beyond the MFV principle and take the alignment hypothesis~\cite{Batell:2017kty}, i.e., the scalar only couples to a single fermion of SM and both the scalar and the Yukawa interactions are simultaneously diagonalizable in a single basis. As emphasized in~\cite{Batell:2017kty}, although the UV origin of the alignment hypothesis remains mysterious, it could give rise to couplings of $S$ predominantly to one flavor in a technically natural way that suppresses lepton flavor violations. Here, we choose $S$ to be dominantly coupled to muons and we refer to this scenario as ``muon-specific''. 

Given that scalar-to-quark coupling is absent and its electron coupling is suppressed in lepton-specific model (or absent in muon-specific model), the effective way to probe the leptonphilic dark scalar is via accelerator experiments with muon beams~\footnote{To be more precise, it is an anti-muon beam generated through proton beam. In the rest of the paper  we indicate both $\mu^-$ and $\mu^+$ as ``muon'' or $\mu$ for simplicity.}. Such experiments have been proposed at Fermilab (FNAL-$\mu$~\cite{Chen:2017awl}, M$^3$~\cite{Kahn:2018cqs}), and CERN (NA64-$\mu$~\cite{Chen:2017awl,Chen:2018vkr}). Alternatively, proposed-proton beam experiments such as SeaQuest~\cite{Berlin:2018pwi}, NA62~\cite{NA62:2017rwk}, or SHiP~\cite{1742-6596-878-1-012014} can also probe the dark scalars via a significant amount of secondary muons or photons. In this paper, we propose a new and alternative way to   prove such a model, making use of  secondary muons generated in electron beam-dump facilities. Such secondary muons have much larger energy and spatial spread in comparison to muon beams originated by protons. Nevertheless, they are abundantly produced and this search can be performed with a minimal upgrade of existing or proposed electron beam-dump experiments.
The novel approach proposed in this paper exploits the fact that, when a high-energy electron impinges on a fixed target, a large variety of secondary particles is produced. This significantly extends the number of physics cases accessible to electron beam dump experiments.

The paper is organized as follows. Section~\ref{sec:sprod} describes in general the muon scalar production on a fixed target. Section~\ref{sec:setup} describes the main properties of electron beam-dump experiments searching for $S$, including: secondary muons production;   scalar particle emission, propagation and decay; and, finally,   decay products detection. Section~\ref{sec:reach_procedure} describes the Montecarlo-based numerical procedure we developed to evaluate the sensitivity of electron beam-dump experiments. Section~\ref{sec:results} presents  results for  E137~\cite{PhysRevD.38.3375} and BDX~\cite{BDX} experiments and, finally,   Sec.~\ref{sec:complementary} discusses limits obtained by complementary  probes  relevant for the parameter space explored in this paper. Appendixes A and B report technical details of the Montecarlo calculation.

\section{\label{sec:sprod} S production by muons and subsequent decay}

\begin{figure*}[t]
\includegraphics[width=0.85\textwidth]{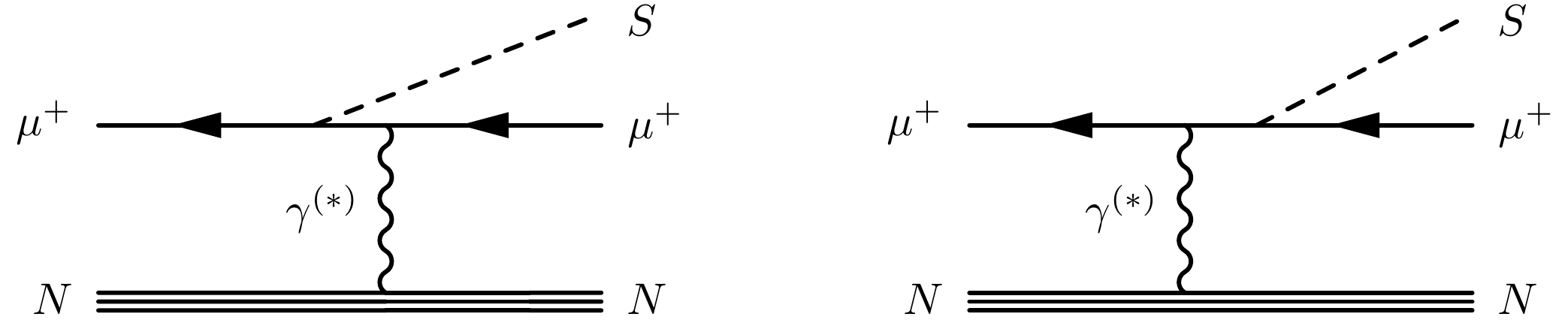}
\caption{\label{fig:diag} Feynmann diagrams for the radiative dark scalar production by an impinging muon on a nucleus $N$. }
\end{figure*}

The main process responsible for scalar emission by an impinging muon on a fixed target is the so-called ``radiative'' production, i.e., the reaction $\mu+N \rightarrow\mu+N+S$, with $N$ being a target nucleus. The corresponding Feynman diagrams are reported in Fig.~\ref{fig:diag}. In case of large muon energy, compared to the muon and to the scalar mass, the corresponding cross-section can be computed at first order using the improved Weizsacker-Williams approximation~\cite{Chen:2017awl}. The differential cross-section, integrated over all the emission angles, reads~\cite{Liu:2016mqv} 
\begin{equation}\label{eq:dsdx}
\frac{d\sigma}{dx} \simeq \frac{g^2_\mu \alpha^2}{12\pi}\chi \beta_\mu \beta_S \frac{x^3[m^2_\mu(3 x^2-4x+4)+2m^2_S(1 - x)]}{[m^2_S (1-x)+m^2_\mu x^2]^2} \; ,
\end{equation}
where $x$ is the scalar energy to muon energy ratio $E_S/E_\mu$, $\beta_\mu= \sqrt{1 - m^2_\mu/E^2_\mu}\simeq 1$ is the muon boost factor, $\beta_S = \sqrt{1 - m^2_S/E^2_S}$ is the scalar boost factor, and $\alpha$ is the fine-structure constant. The effective photon flux $\chi$ is
\begin{equation}
\chi = \int_{t_{\mathrm{min}}}^{t_{\mathrm{max}}} dt 
\frac{t-t_{\mathrm{min}}}{t^2} G_2(t) \; \; ,
\end{equation}
where $-t$ is the square of the four-momentum transferred from the initial to the final state nucleus, $t_\mathrm{min}$ and $t_\mathrm{max}$ are the kinematic limits on $t$, and $G_2(t)$ is the combined atomic and nuclear form factor (see App.~\ref{app:calchep}). The energy distribution of the emitted scalar follows directly from the previous expressions. In case of a light scalar ($m_S \ll m_\mu$), the $x$ distribution is concentrated in the low-$x$ region, with a maximum value at $x_\mathrm{max} \simeq 1.4 \, m_S / m_\mu$. For a heavy scalar, instead, the outgoing $S$ takes more significant portion of the muon energy and the corresponding $x$ distribution is peaked close to $x_{\rm peak}\simeq 1- m_S/E_\mu$. The kinematic of the produced $S$ is strongly peaked in the forward direction, with a typical emission angle \beq
\theta_S \simeq  \sqrt{\frac{m_S^2}{E_\mu^2}\frac{1-\bar{x}}{\bar{x}}+\frac{m_\mu^2}{E_\mu^2}} \eeq where $\bar{x}$ is the typical $x$ value discussed before.

Under our simplified dark scalar model (Eq.~\ref{eq:scalarL}), a light scalar $S$ only decays to pairs of  photons or leptons once the decay channel is kinetically accessible. The total decay width is given by
\begin{align}
\Gamma_S =&\sum_{\ell = e, \mu, \tau} \f{g_\ell^2 m_S}{8\pi} \left(1-\f{4m_\ell^2}{m_S^2}\right)^{3/2}\nonumber\\
+&\f{\alpha^2 m_S^3}{256\pi^3} \left|\sum_{\ell = e, \mu, \tau}\f{g_\ell}{m_\ell} F_{1/2}\left(\frac{4 m_\ell^2}{m_S^2}\right)\right|^2,
\label{eq:width}
\end{align}
where $F_{1/2} (\tau)$ is the fermionic loop-function for on-shell scalar and photons. It is given by
\beq
F_{1/2} (\tau)=\left \{
  \begin{tabular}{l}
     $2 \tau \left[1+\left(1-\tau\right) \left(\arcsin \tau^{-1/2}\right)^2\right]$ \quad $\tau \geq 1$\\\\
      $2\tau \left[1-\f{1-\tau}{4}\left(-i \pi +\ln\f{1+\sqrt{1-\tau}}{1-\sqrt{1-\tau}}\right)^2\right]$ \quad $\tau <1$
  \end{tabular}
\right.
\eeq

In this paper, we primarily focus on  $2 m_e < m_S < 2 m_\mu$ case, resulting in a dominant decay into an $e^+ e^-$ pair for the lepton-specific model or to a $\gamma \gamma$ pair for the muon-specific case. This is motivated by the fact that the beam dump experiments here considered have maximum sensitivity to these channels.  
The corresponding partial widths are
\begin{align}
\Gamma_{e^+e^-}= & \frac{g^2_e m_S}{8\pi}\left(1-\frac{4m^2_e}{m^2_S}\right)^{3/2} \label{eq:gamma_ee}
=
 \frac{g^2_\mu m^2_e  m_S}{8\pi m^2_\mu}\left(1-\frac{4m^2_e}{m^2_S}\right)^{3/2} 
\\
\Gamma_{\gamma\gamma}= & \frac{\alpha^2 m^3_S}{256 \pi^3}
\left| 
\frac{g_\mu}{m_\mu}F_{1/2}
\left(
\frac{4m^2_\mu}{m^2_S}
\right) 
\right|^2 \;\;,
\end{align}
where in the second equality of Eq.~\eqref{eq:gamma_ee} we used the identity $g_e / m_e = g_\mu / m_\mu$ valid for the lepton-specific model. We note that, for same values of model parameters $m_S$ and $g_\mu$, $\Gamma_{e^+e^-} \gg \Gamma_{\gamma\gamma}$, resulting in longer-lived $S$ particles for the muon-specific model than for the lepton-specific one. 

Finally, we observe that, for the case $m_S > 2 m_\mu$, $S$ decays always to a $\mu^+ \mu^-$ pair, with larger decay width
\begin{equation}
\Gamma_{\mu^+\mu^-}=  \frac{g^2_\mu m_S}{8\pi}\left(1-\frac{4m^2_\mu}{m^2_S}\right)^{3/2} \; \; .
\end{equation}
This results in a reduced sensitivity for a typical beam-dump experiment, since the large decay width required to cope with the experimental acceptance (see next Sec.) results to a $g_\mu$ value, and hence to a production cross-section, too small to have an appreciable production yield.
\section{\label{sec:setup} Searching for $S$ with secondary muons at electron beam dumps}

\begin{figure*} 
\includegraphics[width=.85\textwidth]{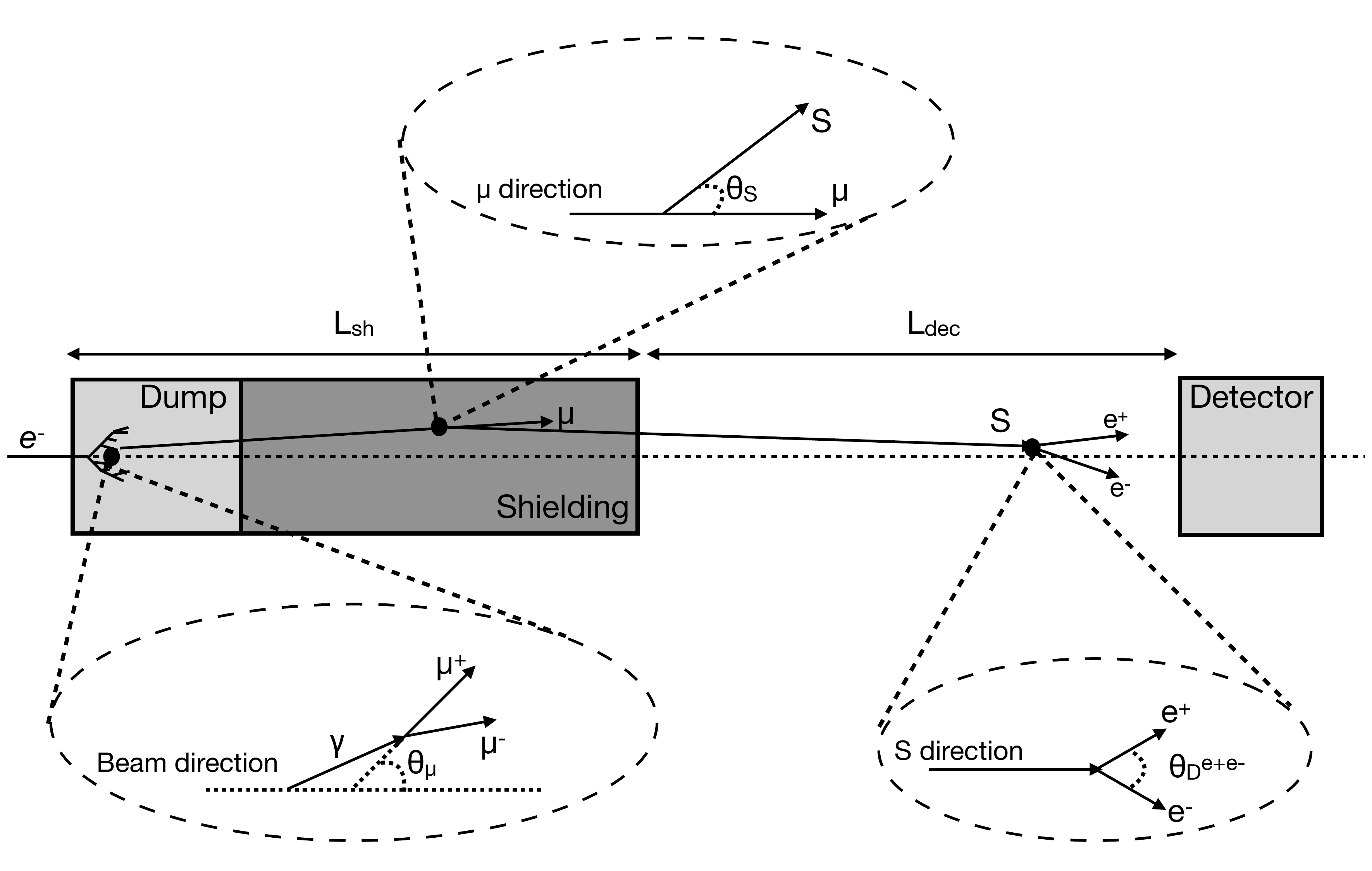}
\caption{\label{fig:decayScheme} Typical setup of an ``E137-like'' electron beam-dump experiment for visible decay $S$ search through secondary muons. $L_{\rm sh}$ is the total length of target and shielding, while $L_{\rm dec}$ is the length of the downstream decay region, preceding the detector. The three insets show schematically the production of a $\mu^+ \mu^-$ pair by a photon in the dump, the production of $S$ by a secondary muon, and the $S$ decay to an $e^+ e^-$ pair. For a ``BDX-like'' experiment, instead, the shielding extends to the detector position, and the decay region overlaps with the detector volume.}
\end{figure*}

The typical setup of an electron beam dump  experiment searching for muonphilic scalar particles is shown in Fig.~\ref{fig:decayScheme}.
Scalar particles are produced by the primary electron beam impinging on the fixed target through a tertiary process involving secondary muons. These propagate in the target and in the surrounding materials, undergoing energy loss and multiple scattering, and radiatively emit the scalar particles. The resulting $S$ kinematic distribution, including the production vertex, is thus given by the convolution of the muon distribution, altered by energy loss and multiple scattering, and the differential production cross-section.

After being produced, $S$ particles propagate straight, until decaying to an $e^+ e^-$ or $\gamma \gamma$ pair. These particles are measured by a detector placed behind the beam-dump. A sizable amount of shielding material is placed between the dump and the detector to range out all other particles produced by the primary beam. Two detection setups are hence possible. In the first case (E137-like), the detector is sensitive to $S$ decay products ($e^+$, $e^-$, or $\gamma$) produced  in a free decay region downstream of  the shielding. In the second scenario (BDX-like), the shielding extends up to the detector location, and the decay region overlaps with the detector volume.


\subsection{Muons production in $e^-$ beam dumps}

Production of high-energy muons by an impinging $e^-$ on a heavy thick target predominantly happens via two different mechanisms, the radiative emission of a $\mu^+ \mu^-$ pair and the decay-in-flight of photo-produced pions and kaons. The contribution of muons produced by the decay-at-rest of $\pi$'s and $K$'s can be neglected due to the very low emission energy, typically lower than the scalar production and detection thresholds.

Muons with energy of the order of the beam energy are mainly produced by pair emission, predominately happening via a two-step process~\cite{Cox:1968cq}. First, an $e^-$ radiate a  bremsstrahlung photon in a nucleuos field, then the photon produces the $\mu^+ \mu^-$ pair in a close-by nucleus. The direct production process through a virtual photon exchange, $e^- N \rightarrow e^- N \gamma^* \rightarrow e^- N \mu^+ \mu^-$, is instead negligible, due to the much lower flux of virtual photons. The kinematics of produced muons is strongly peaked in the forward direction, with the typical magnitude of transverse momentum being $p_{{\rm T}, \mu} \simeq m_\mu$. 
Being the pair-production on nuclei a coherent process, the yield of muons produced by the neutral component of the electro-magnetic shower through this reaction is almost independent on the target material (see~\cite{NELSON1968293}. Eq.~9).

In the low energy range, instead, the dominant production mechanism is the decay-in-flight of photo-produced $\pi$'s and $K$'s. The kinematics of emitted muons is isotropic in the hadron rest frame, and shows a forward peak in the laboratory frame whose width depends on the boost factor of the decaying parents. Since $\pi$'s and $K$'s photo-production on nuclei scales roughly as the atomic number, the corresponding muon yield shows a significant $1/Z$ material dependence, $Z$ being the atomic number of the target.

\begin{figure}[tpb]
\includegraphics[width=.5\textwidth]{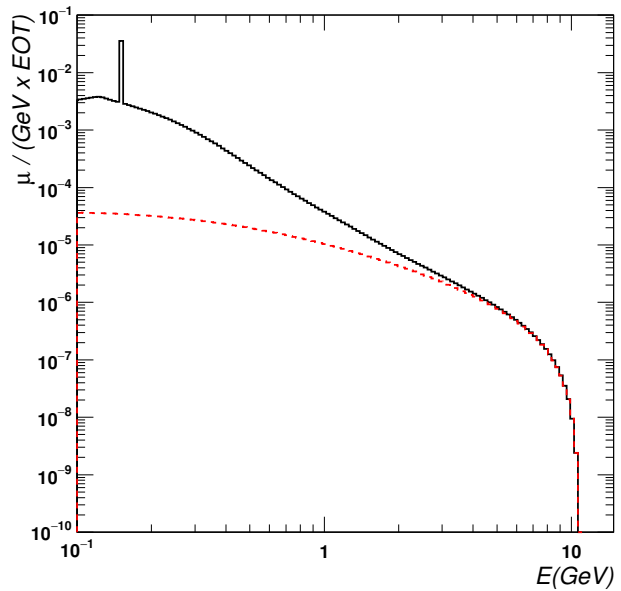}
\caption{\label{fig:muFlux} (Colors online). The differential muons yield per EOT for  11 GeV $e^-$ beam a impinging on a thick aluminum and water target, as a function of the muon kinetic energy. The continuous black curve refers to all produced muons, while the dashed red curve refers only to pair-produced muons. The peak in the full distribution at $E=152$ MeV is due to the kaon decay-at-rest process, $K \rightarrow \mu \nu_\mu$.}
\end{figure}

Figure~\ref{fig:muFlux} shows the kinetic energy distribution of muons produced by an 11 GeV $e^-$ beam impinging on a thick aluminum and water target, comparing the full yield to the contribution of photo-produced muons only.
This result was obtained from a \texttt{FLUKA}-based~\cite{BOHLEN2014211,Ferrari:2005zk} Montecarlo simulation, where we implemented the specific target geometry and composition foreseen in the BDX experiment (see Sec.~\ref{sec:BDX}).
The overall muon yield per electron on target (EOT), in the energy range $0.5 - 11$ GeV ($2 - 11$ GeV)  is $7.65 \cdot10^{-5}$ muons/EOT ($0.91 \cdot10^{-5}$ muons/EOT).
For a primary $e^-$ beam current of about 10 $\mu$A,  this corresponds to a muon rate of $\mathcal{O}(10^{9}-10^{10})$ muons/second, depending on the considered energy range. This large flux demonstrates the potential of secondary muons at a multi-GeV electron beam dump facility, in particular when compared to the typical intensity of existing muon beams with similar energies (the Fermilab accelerator complex, for example, can deliver a muon beam of about $10^7$ muons/second \cite{Chapelain:2017syu}).

\begin{figure*}[tpb]
\includegraphics[width=.45\textwidth]{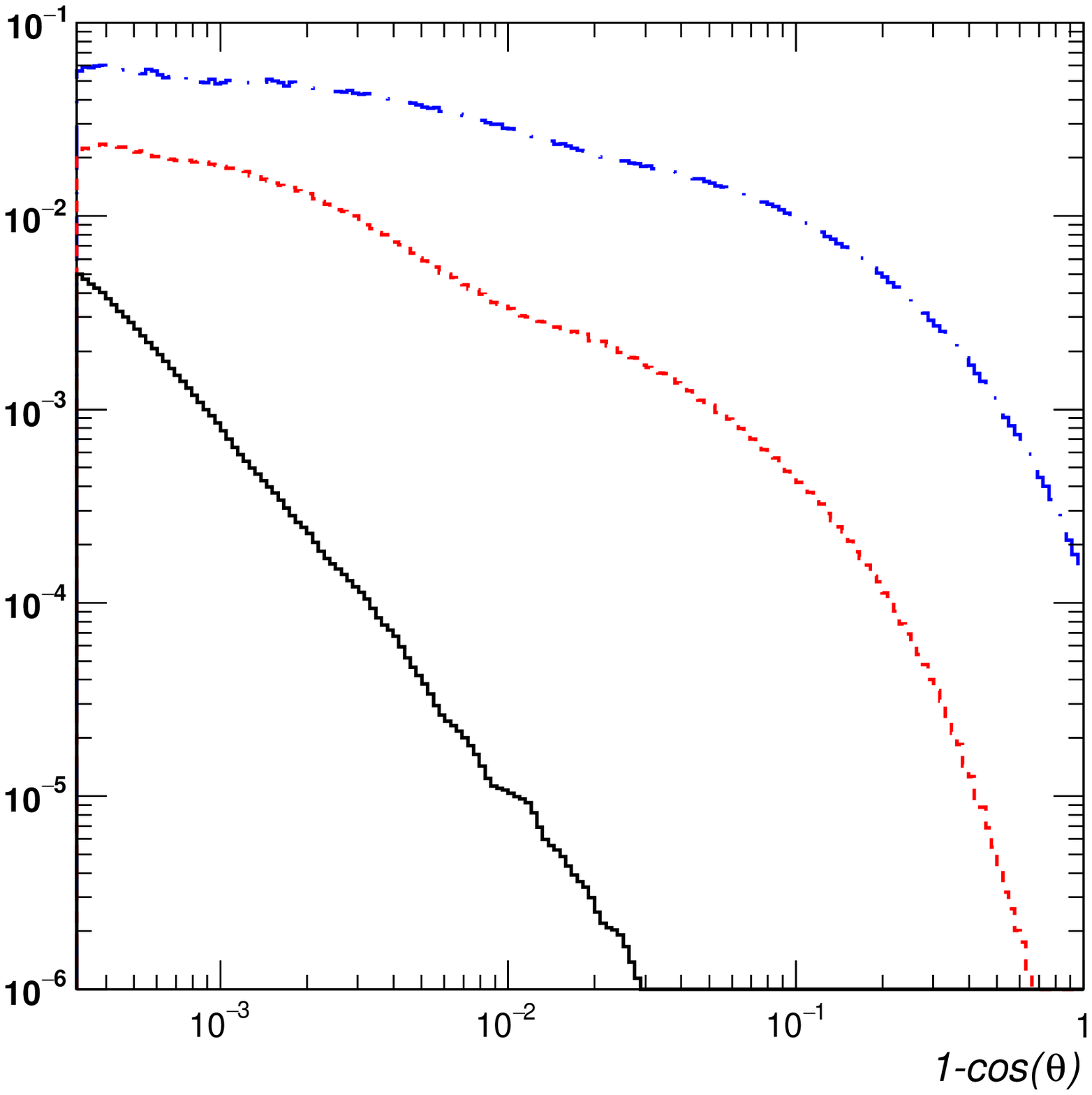}
\includegraphics[width=.45\textwidth]{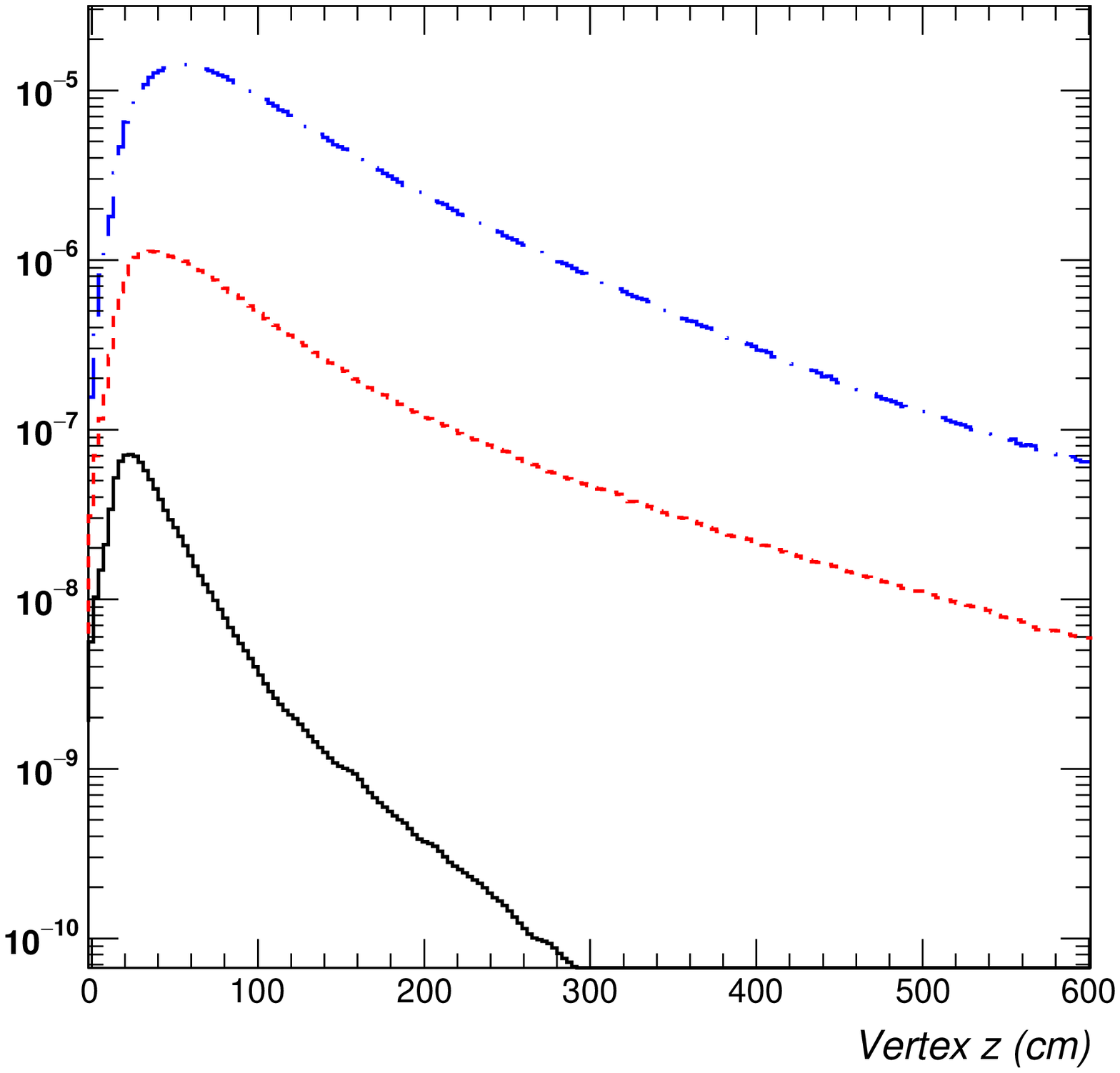}
\caption{\label{fig:muKin} Kinematic distribution of muons produced in a thick aluminum and water target hit by 11 GeV electrons. Left: angular distribution, right: longitudinal vertex coordinate distribution (the beam impinges on the thick target at $z=0$). The continuous black, dashed red, dot-dashed blue curves refer, respectively, to 2 GeV, 1 GeV, and 0.5 GeV muons. The normalization of each curve is proportional to the total muon yield at the corresponding energy.}
\end{figure*}

Figure~\ref{fig:muKin} shows the kinematic distribution of produced muons, in terms of the emission angle and vertex longitudinal coordinate $z$ (for the latter, $z=0$ corresponds to the target front face.). Higher-energy muons are mostly produced in the forward direction, in the first target radiation lengths, while for lower energies the angular distribution is wider, and particles are produced over the full target length.

\subsection{$S$ production}

Muons produced in the thick target penetrate deeply in it and surrounding materials, losing energy mainly through ionization~\cite{Patrignani:2016xqp}. While traveling, they may produce a scalar particle through radiative emission on a nucleus.

In the simplest case where the target sizes are larger than the average muons range, and considering an uniform material, the total $S$ production yield is given by

\begin{equation}\label{eq-prod}
N_S= \frac{N_A}{A} \rho \int_{E_{\min}}^{E_0} dE_\mu \,\, T_\mu(E_\mu)\,\sigma(E_\mu) \; ,
\end{equation}
where $A$ and $\rho$, are, respectively, the target material atomic mass and mass density, $E_0$ is the primary electron-beam energy, $N_A$ is Avogadro's number, $\sigma(E_\mu)$ is the energy-dependent $S$ production cross-section, and $E_{\min}\simeq m_S$ is the minimal muon kinetic energy required to produce a scalar with mass $m_S$ through radiative emission on a nucleus.
Finally, $T_\mu(E_\mu)$ is the muons differential track-length distribution in the target, defined as the integral over the target volume of the differential muons fluence $\Phi_\mu(E_\mu)$, corresponding to the density of particle tracks in the volume \cite{Chilton}. Intuitively, the quantity $T_\mu(E_\mu)dE_\mu$ represents the total path length in the thick target of muons with kinetic energy in the interval between $E_\mu$ and $E_\mu+dE_\mu$.
At the first order, neglecting multiple scattering and considering a constant stopping power $\left< dE/dx\right>$ ($\sim \rho \cdot 2\, {\rm MeV}\,{\rm g}^{-1}\,{\rm cm}^{-2}$), $T(E_\mu)$ is given by
\begin{equation}
T_\mu(E_\mu) = \frac{\int_{E_\mu}^{E_0} n_\mu(E)dE}{\left<dE/dx\right>} \; \; ,
\end{equation}
where $n_\mu(E)$ is the differential yield of muons in the target. From previous considerations about muon production in the target, and given the shape of $n_\mu(E)$, it follows that $T(E_\mu)$ depends on the target material as $(\rho Z)^{-1}$ in the low $E_\mu$ region, and as $\rho^{-1}\log{Z}$ at higher muon energy. 
For a scalar in the mass range here considered most of the contribution to the integral in Eq.~\ref{eq-prod} comes from the low $E_\mu$ region ($E_\mu \lesssim$ 1 GeV). Therefore, considering the $Z^2$ dependence of $\sigma(E_\mu)$~\cite{Chen:2017awl}, it follows that the total scalar yield depends weekly on target material. 

The kinematic distribution of scalar particles, including the production vertex, replicates the distribution of secondary muons in the beam dump, convolved with the differential production cross-section and distorted by the material-dependent mean free path for the $\mu N \rightarrow \mu N S$ process. In particular, the energy distribution of scalar particles (integrated over the full angular range) reads
\begin{equation}\label{eq-diff}
\frac{dN_S}{dE_S}= \frac{N_A}{A} \rho \int_{E_{min}}^{E_0} dE_\mu \,\, T(E_\mu)\,\frac{d\sigma}{dE_S} \; ,
\end{equation}
with $\frac{d\sigma}{dE_S}$ given by Eq.~\ref{eq:dsdx}. From the previous discussion about the shape of $\frac{d\sigma}{dE_S}$, it follows that for large $m_S$ values, where $\frac{d\sigma}{dE_S}$ is peaked at $E_S \simeq E_\mu$, $\frac{dN_S}{dE_S} \propto T_\mu(E_S)$. For lower mass values, instead, the broadening induced by $\frac{d\sigma}{dE_S}$ is larger. Finally, we note that the shape of $\frac{dN_S}{dE_S}$ depends significantly on the target material. For example, the most energetic part of the spectrum, for $E_S \simeq E_0$, is due to the high-energy part of $T(E_\mu)$, and thus exhibits a $Z\log Z$ dependence.

Previous considerations were derived for a simplified geometry of a large and homogeneous target. In a more realistic case, the target may be non-homogeneous, and its size may be smaller than the range of muons, requiring a torough description of  surrounding materials. The evaluation of the total $S$ yield and of the corresponding kinematics thus requires to compute separately $T(E_\mu)$ in each geometry element, by tracking muons in the full experimental setup. For a realistic and accurate evaluation of real experimental setups, we performed this calculation numerically through an ad-hoc Montecarlo simulation, as described in Sec.~\ref{sec:reach_procedure}.

\subsection{Signal yield in the detector}

After being produced, $S$ propagates straight with a differential decay probability per unit path given by
\begin{equation}
\frac{dP}{dl}=\frac{1}{\lambda}e^{-l/\lambda} \; ,
\end{equation}
where
\beq 
\lambda = \frac{E_S}{m_S}\frac{1}{\Gamma_S}
\eeq is the energy-dependent $S$ decay length.

Particles from the $S$ decay are emitted on a cone with typical aperture \beq \theta_D\simeq \frac{m_S}{E_S} \eeq with respect to the $S$ direction. The total signal yield is obtained combining the $S$ angular and vertex distribution with the decay kinematics, and projecting the result over the geometrical acceptance of the detector. 
The latter can be roughly determined as the product of a longitudinal factor $\varepsilon_L$ depending on the shielding $L_{\rm sh}$ and decay region $L_{\rm dec}$ length and a transverse factor $\varepsilon_T$ related to the detector transverse area $A_T$ (see Fig.~\ref{fig:decayScheme}). For a single scalar particle produced with momentum $\vec{p}_S$ and longitudinal vertex coordinate $z_S$, the longitudinal factor reads
\begin{equation}
P_L  \sim  e^{(z_S-L_{\rm sh})/\lambda}  (1-e^{-L_{\rm dec}/\lambda}) \; \; .
\end{equation}
The longitudinal acceptance $\varepsilon_L$ is obtained convolving $P_L$ with the scalar particles kinematic distribution.
The transverse factor, instead, reads
\begin{equation}
\varepsilon_T \sim \f{A_T}{(\theta^{\rm rms}_S (L_{\rm sh}+L_{\rm dec}-\overline{z}_S) \oplus \theta_D \overline{L}_D)^2} \; ,
\end{equation}
with $\theta^{\rm rms}_S$ being the width of the scalar angular distribution, $\theta_D$ the typical opening angle between the $e^+e^-$ or $\gamma \gamma$ pair from $S$ decay, $\overline{z}_S$ the average $S$ production vertex, and $\overline{L}_D$ the average distance between the $S$ decay point and the detector - in case of a BDX-like setup, with $S$ decaying within the detector volume, $\overline{L}_D$=0.

\section{\label{sec:reach_procedure} Reach Evaluation}

To evaluate the BDX reach and to determine the exclusion limit set by  E137 null result for the dark scalar models, one has to determine, for a given combination of the model parameters ($m_S$ and $g_\mu$), the expected number of signal events within the detector acceptance. We performed the calculation of the signal yield numerically, decoupling the evaluation of $S$ particles production from the subsequent propagation and decay as follows. 

First, we pre-computed the yield of muons produced by the primary electron beam through an optimized simulation using \texttt{FLUKA 2011.2x.3}. In the simulation, we implemented the description of each experimental setup (geometry and materials), including the thick target, the surrounding materials, and the subsequent shielding. We applied different biasing techniques, namely cross-section enhancement and leading-particle biasing for electromagnetic showers, to enhance forward muon production. The obtained result is a list of produced muons momenta, vertexes and statistical weights, together with the total yield per EOT. 

Each muon was then individually tracked through a \texttt{Geant4}-based simulation~\cite{AGOSTINELLI2003250}, implementing the same geometry as implemented in \texttt{FLUKA}. 
We modified the \texttt{Geant4 4.10.02} source code to include the new scalar particle $S$ and the radiative $S$ emission process (more details are given in App.~\ref{app:g4}). When a scalar particle is produced, the corresponding four momentum and production vertex are saved to an output file, and the particle is not further tracked. The output of this second step is a list of produced $S$ particles, together with the total normalization.

This result was finally used as input for a custom Montecarlo code that handles the $S$ propagation and subsequent decay to an $e^+e^-$ or $\gamma \gamma$ pair and computes the experimental acceptance  of a detector placed downstream of the dump, $\varepsilon^\mathrm{det}$, including geometrical and detection threshold effects.

The overall normalization and the signal yield dependence on model parameters were handled as follows. The first computation step only considers $S$ production, and the corresponding result depends on model parameters through the total number of produced $S$ particles per EOT, $N^0_S \propto g^2_\mu$. The corresponding kinematics, instead, does not depend on $g_\mu$. Therefore, only the mass of the scalar, selected at the beginning of the simulation, was left as free parameter while the coupling $g_\mu$ was fixed at a conventional value of $g_\mu^0\equiv 3.87\cdot10^{-4}$. 
The detection acceptance computed in the second step, instead, depends critically on the $g_\mu$ value, affecting the $S$ decay length. Therefore, the evaluation of $\varepsilon^\mathrm{det}$ has been performed as a function of this parameter.
For a given combination of $m_S$ and $g_\mu$, the total signal yield in the detector per EOT thus reads
\begin{equation}
N_S(m_S,g_\mu) = N^0_S(m_S) \cdot \varepsilon^\mathrm{det}(m_S,g_\mu) \cdot \left(\frac{g_\mu}{g^0_\mu}\right)^2 \; \; .
\end{equation}

We found this multi-step method more effective than performing a single \texttt{Geant4} simulation, including both the muon production by the developing electromagnetic shower and the propagation and decay of $S$, since it saves a significant computation time in the evaluation of the sensitivity of a beam-dump experiment as a function of the $S$ mass and coupling. Indeed, for a given experimental setup, the muon yield estimation is performed only once. Then, for a given mass value, the $S$ production is evaluated, and only the computation of detector acceptance is repeated for different values of the scalar coupling.

\section{\label{sec:results} Results}

In this section, we present the exclusion limit in the $m_S$ vs $g_\mu$ parameters space obtained from the E137 experiment, and the expected sensitivity for the
and BDX experiment, for both the lepton-specific and the muon-specific models. Results are summarized in Fig.~\ref{fig-result}, reporting also published limits from other experiments (see Figure caption for further details).

\subsection{E137 exclusion limits}

\begin{figure}[t] 
\includegraphics[width=.48\textwidth]{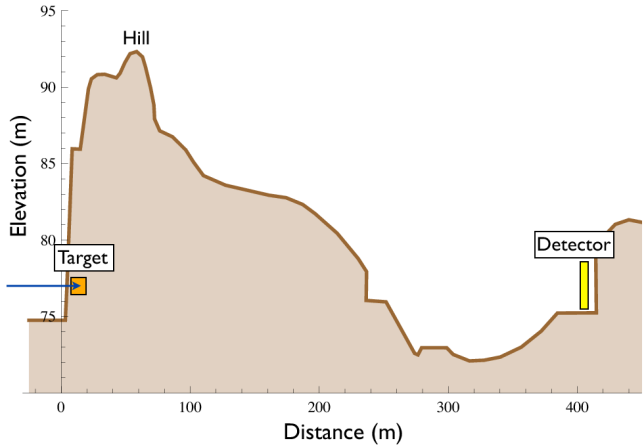}
\caption{\label{fig:E137geometry} (Colors online). Layout of the E137 experiment, showing distance between the beam dump (``target'') and the detector. Figure reproduced from Ref.~\cite{PhysRevLett.113.171802} with authorization from authors.
}
\end{figure}

The E137 experiment searched for long-lived neutral objects produced in the electromagnetic shower initiated by 20 GeV electrons in the SLAC beam dump.
Since muons penetrate deeply in the thick target and in the surrounding materials, a precise description of the experimental setup geometry and materials is required. The E137 beam dump was a 720 cm long aluminum tank completely filled with water~\cite{PhysRevA.29.2110,Walz:1967mg}, with an external diameter of 152 cm. The thickness of the aluminum vessel was 0.9525 cm, a part from the hemispherical front window, with a reduced thickness of 0.475 cm. The vessel radius was 75 cm. The underground hall hosting the dump was made by concrete walls, with thickness of about 90 cm. The distance between the dump end and the front wall was 160 cm. To protect the concrete wall from low-energy radiation escaping the dump, a 56 cm 
long iron block was added. The transverse size of the iron block, determined from the original engineering drawings, was comparable to that of the beam dump.

Scalar particles produced in the thick target would have to penetrate 179 m of earth shielding and decay in the 204 m region downstream of the shield. 
We assumed a uniform dirt density of $\rho_{\rm dirt} = 1.7\,{\rm g}/{\rm cm}^{3}$, a value that seems to us reasonable considering what reported in Ref.~\cite{Seryi:453643} and given the average depth of soil of about 10 m for the terrain following the underground hall~\cite{PhysRevD.38.3375}.
To evaluate the effect of this parameter on the final result, we repeated the E137 reach computation changing it by $\pm 10\%$, obtaining  a negligible variation of order $\sim  1 \% $.

The E137 detector was an 8-radiation length electromagnetic calorimeter made by a sandwich of plastic scintillator paddles and iron (or aluminum) converters. Multi-wire proportional chambers provided an accurate angular resolution, essential to keep the cosmic background to a negligible level. A total charge of $\sim$ 30~C was dumped during the live-time of the experiment in two slightly different experimental setups: in the first run (accumulated charge $\simeq$~10~Coulomb), the detector had a transverse size of 2$\times$3 m$^2$, while in the second run this was 3$\times$3 m$^2$.  

The original data analysis searched for axion-like particles decaying in $e^+ e^-$ pairs, requiring a deposited energy in the calorimeter larger than 1 GeV with a track pointing to the beam dump production vertex. The absence of any signal provided stringent limits on axions or photinos.

To derive the E137 exclusion limits on dark scalar production, we used the Montecarlo-based numerical approach described above.
The experimental acceptance was evaluated separately for the two E137 runs and combined with proper weights to account for the different accumulated charges. In the calculation, we employed the same selection cuts used in the original analysis, requiring that
the energy of the impinging particle is larger than 1 GeV and that the corresponding angle of impact on the detector, measured with respect to the primary beam axis, is smaller than 30 mrad.
We found that both particles from $S$ decay hit the detector in a non-negligible fraction of events. In these cases, we applied previous selection cuts respectively considering the sum of the two energies to be greater than 1 GeV and the energy-averaged impinging angle to be less than 30 mrad.
Based on the null observation reported by E137, we derived the exclusion contour considering a $95\%$ C.L. upper limit of 3 events.

\subsection{BDX sensitivity}\label{sec:BDX}

\begin{figure}[t] 
\includegraphics[width=.5\textwidth]{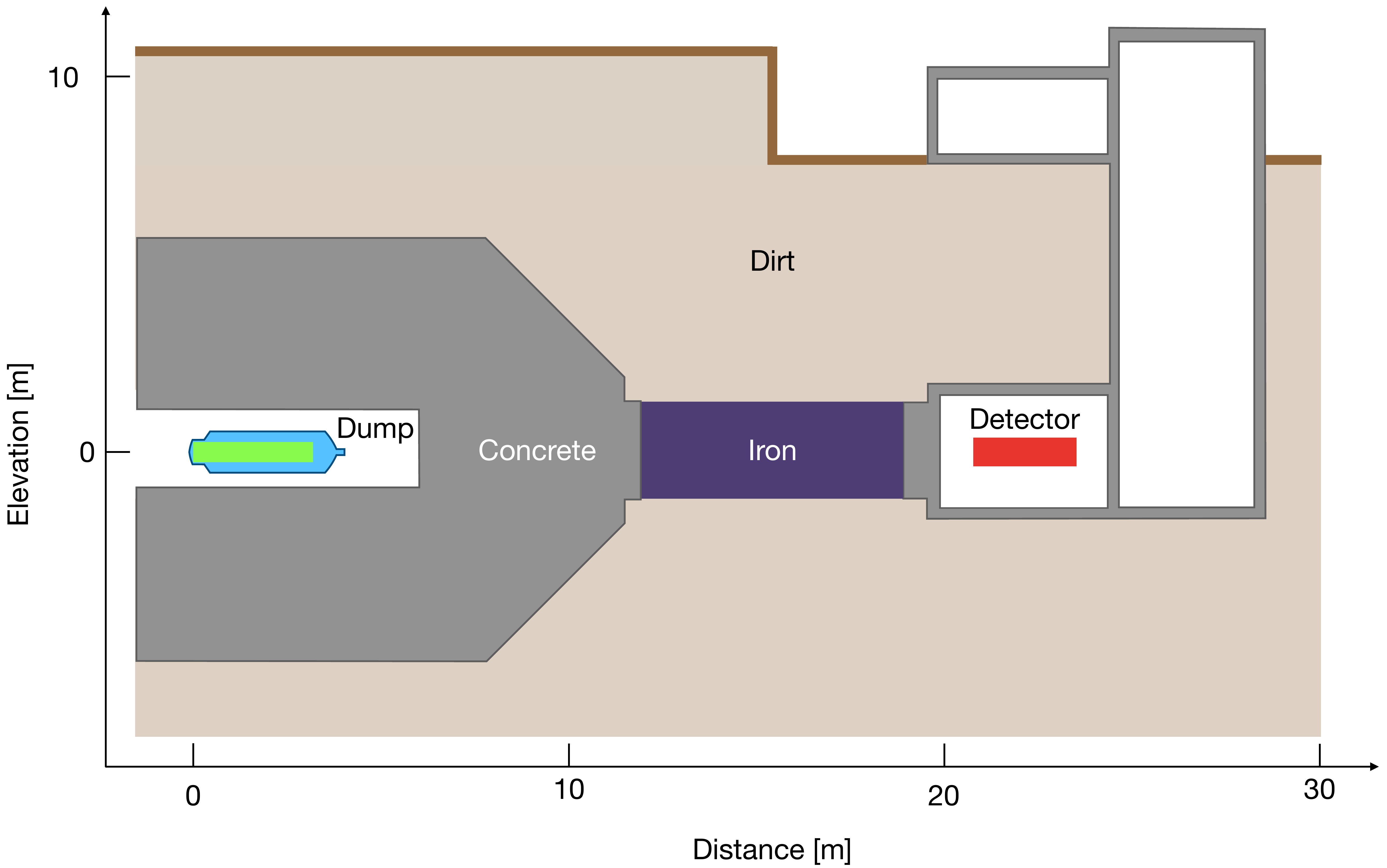}
\caption{\label{fig:BDXgeometry} (Colors online). Implementation of the BDX experimental setup in \texttt{FLUKA}. From left to right, the Hall-A aluminum-water beam dump (blue-green), the concrete walls (gray), the iron shielding (purple), and the detector (red) in the new experimental hall. The two scales report distances between elements (in meters). The detector is located approximately 20 m downstream the beam dump, 8 m underground.}
\end{figure}

BDX is a planned electron-beam dump experiment at JLab that will improve the E137 sensitivity by order of magnitudes by using the high intensity 11 GeV CEBAF beam~\cite{Freyberger:2015rfv}, running for $\sim$1 year with currents up to $60$ $\mu$A, thus collecting $\simeq 10^{22}$ EOT.
BDX will make use of the Hall-A beam dump~\cite{HallABeamDump}, with the detector placed 20 m downstream in a new, dedicated experimental Hall (see also Fig.~\ref{fig:BDXgeometry}). In this configuration, the experiment will produce $\sim 10^{17}$ GeV-energy muons in the dump, a total yield much larger than what can be obtained in similar fixed-target efforts employing a primary muon beam~\cite{Chen:2017awl}.

The BDX detector consists of a $50\times40\times300$ cm$^3$ electromagnetic calorimeter made by 800 CsI(Tl) scintillating crystals, surrounded by two active veto layers made by plastic scintillator for cosmic backgrounds rejection. An iron and concrete shielding layer will be installed between the existing Hall-A beam dump vault and the detector hall to range out all other particles produced in the target (except neutrinos).

The primary physics goal of BDX is the search for light dark matter particles, produced by the primary electron beam in the target, by measuring their scattering on atomic electrons in the detector, resulting in high-energy electromagnetic shower~\cite{Battaglieri:2016ggd}. In particular, the experiment was designed and optimized considering the physics case of an invisibly-decaying dark photon~\cite{PhysRevD.88.114015}. In case of a negative result, BDX is expected to improve current exclusion limits by about two order of magnitudes.
The experiment is also sensitive to long-lived particles decaying visibly to an $e^+e^-$ or $\gamma \gamma$ pair, if the decay occurs within the detector volume. In this work, we derived the expected BDX sensitivity to a scalar particle considering, conservatively, an expected background contribution of $5$ events, corresponding to a 95$\%$ CL sensitivity of $\sim 6$ signal events.

\begin{figure*}[t] 
\includegraphics[width=.5\textwidth]{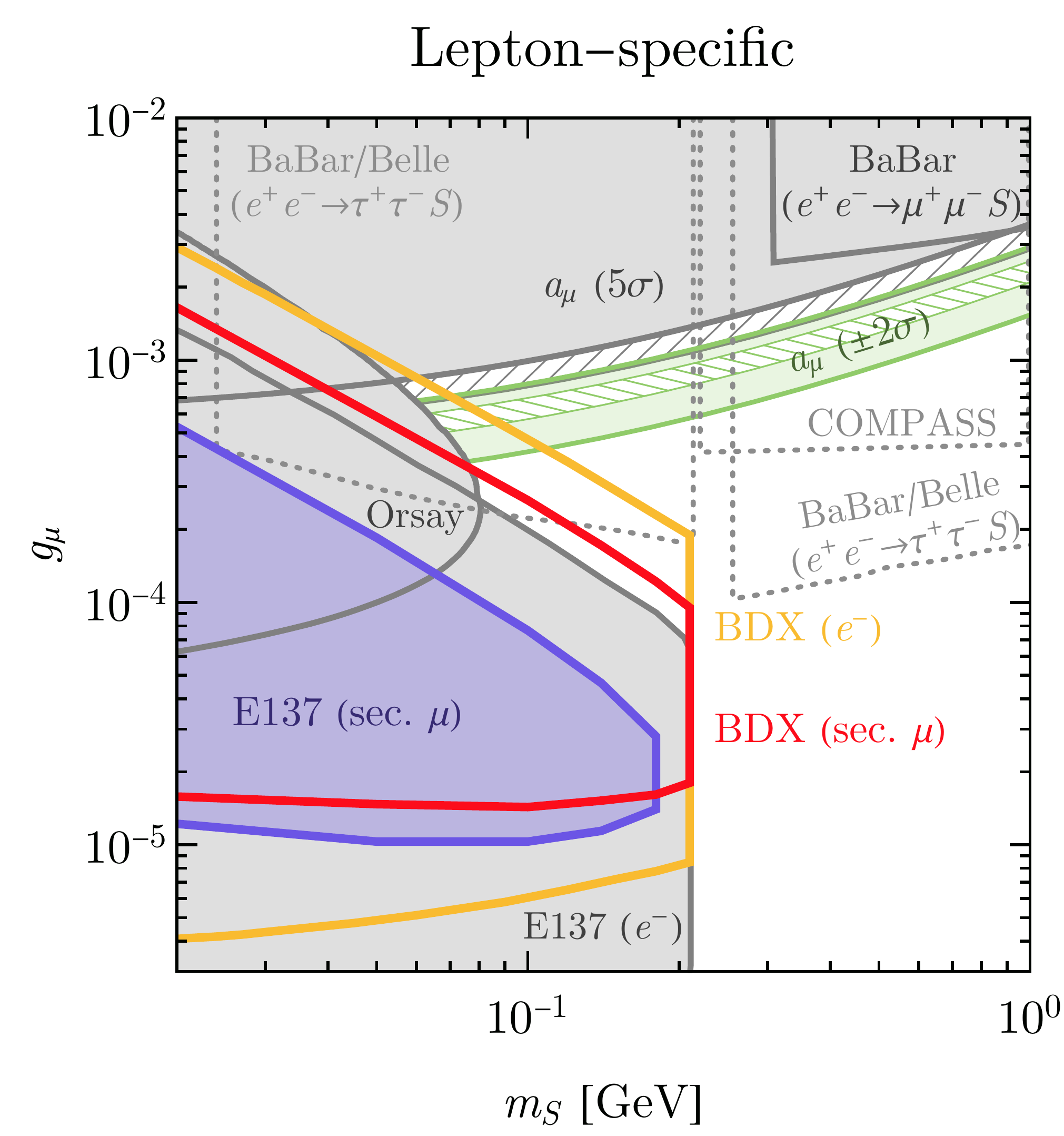}~\includegraphics[width=.5\textwidth]{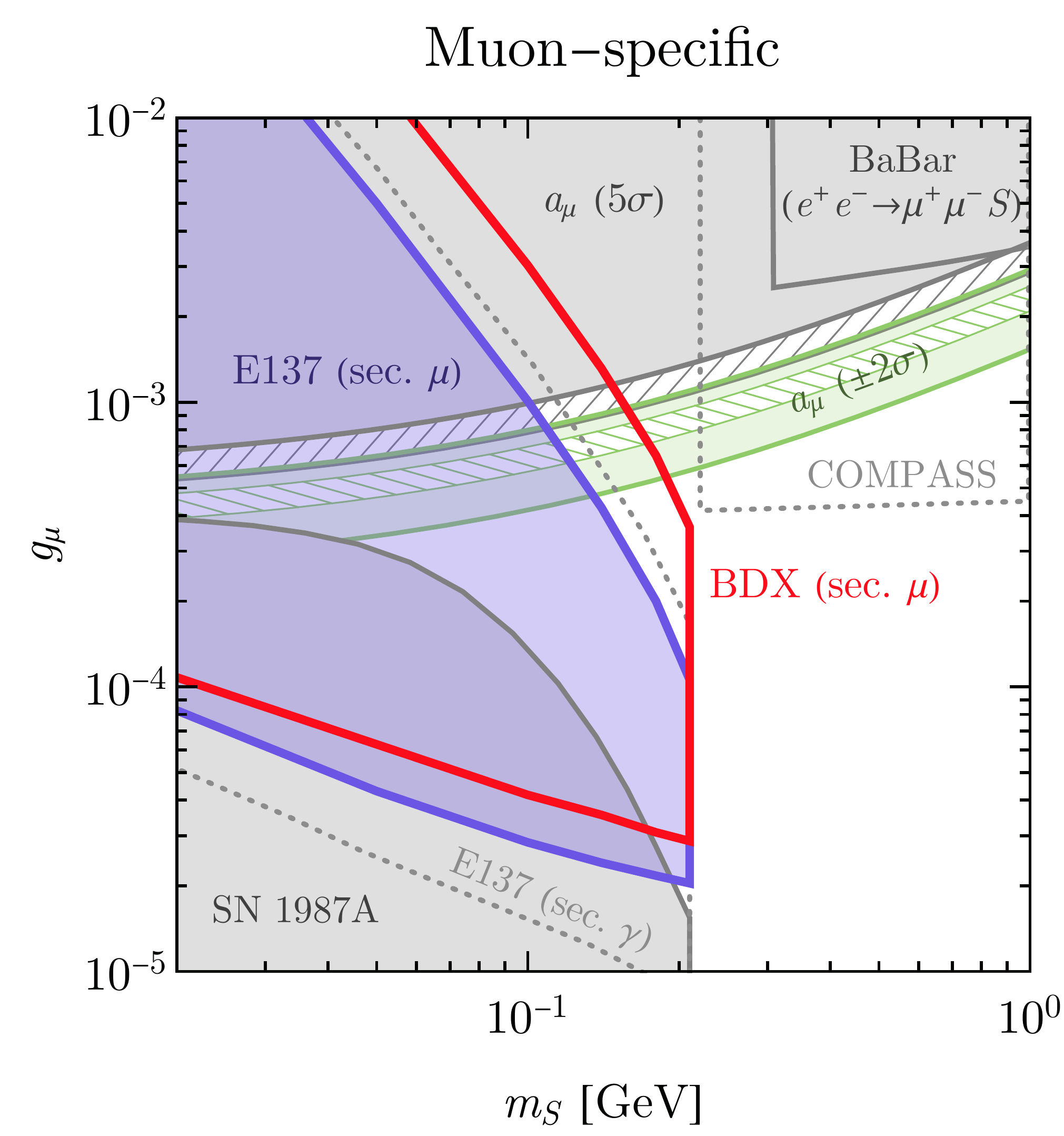}
\caption{\label{fig-result}(Colors online). 95\% CL exclusions and projections on dark scalar searches from secondary muons at E137 and BDX cross the $g_\mu$-$m_S$ plane, along with several other experimental projections and existing constraints. The left and right panel corresponds to the lepton-specific model ($g_e:g_\mu:g_\tau =m_e:m_\mu:m_\tau$) and muon-specific model ($g_\mu \neq 0, g_e =g_\tau =0$). The blue regions represent constraints from secondary muons produced at E137. The red contours represent expected sensitivities from secondary muons produced at BDX. The yellow contour in the left panel represents the expected sensitivity through electron bremsstrahlung at BDX. The green regions show the 2$\sigma$-favored region from current $a_\mu$ experiment~\cite{Olive:2016xmw}. The gray regions are combined exclusions based on existing experiments or observations. For the lepton-specific model, the gray region includes current $a_\mu$ measurement with $5\sigma$ exclusions, BaBar limits based on $e^+ e- \to \mu^+ \mu^- S$ searches~\cite{TheBABAR:2016rlg}, and constraints from electron bremsstrahlung at Orsay and E137~\cite{Batell:2016ove}. For the muon-specific model, the gray region includes the same exclusions from $a_\mu$ and BaBar as the ones for the lepton-specific model. Besides it also includes constraints from SN 1987A (see Sec.~\ref{sec:sn} for more details). The hashed regions in both panel indicate the projected sensitivity of future $a_\mu$ measurements, assuming the current central value stays the same while the experimental and theoretical uncertainties will be improved by a factor of 4 and 2, respectively~\cite{Grange:2015fou, Chapelain:2017syu, Mibe:2010zz, Jegerlehner:2018zrj}. The dotted contour shows expected sensitivity if BaBar or Belle could search for $e^+ e- \to \tau^+ \tau^- S$~\cite{Batell:2016ove} or COMPASS could search for $\mu N \to \mu N S$~\cite{Essig:2010gu} with existing data. Note that we also indicate constraints from the secondary photons at E137~\cite{Dolan:2017osp} as a dotted line given the uncertainties in the analysis as we discussed in Sec.~\ref{sec:complementary-dump}. Projected sensitivities for leptophilic dark scalars produced by future muon-beam experiments or secondary muons/photons at future proton-beam experiments can be found in~\cite{Chen:2017awl,Berlin:2018pwi}.}
\end{figure*}


\section{Complementary Probes}\label{sec:complementary}
\subsection{Anomalous magnetic moments}
One of the most sensitive bounds on light scalars for lepton-specific or muon-specific models is from the measurement of muon anomalous magnetic moment $a_\mu \equiv (g-2)_\mu/2$. Current measured value of $a_\mu$ is larger than the SM prediction by~\cite{Olive:2016xmw}
\begin{equation}
\Delta a _\mu = a_\mu (\text{EXP})-a_\mu(\text{SM}) = (268 \pm 63\pm 43)\times 10^{-11},
\end{equation}
where the first and the second error bars indicate the experimental and theoretical systematical uncertainties, respectively. Introducing a new scalar that couples to muons can uplift the predicted value of $a_\mu$ by
\begin{equation}
\Delta a_\mu (S) = \frac{g_\mu^2}{8\pi^2}\int^1_0 dz \f{(1-z)^2(1+z)}{(1-z)^2+z(m_S/m_\mu)^2},
\end{equation}
and makes it agree with the current measured value. In the $g_\mu$--$m_S$ parameter space, we show the favored regions consistent with the measured value of $a_\mu$ as the shaded green region with label $a_\mu(\pm2\sigma)$. We also report the excluded regions that yield a $5\sigma$-discrepancy with respect to the measured value as the shaded gray region (joint with other exclusions) with label $a_\mu (5\sigma)$. On-going experimental and theoretical developments may reduce the experimental and theoretical systematical uncertainties by a factor of 4 and 2 respectively in near future~\cite{Grange:2015fou, Chapelain:2017syu, Mibe:2010zz, Jegerlehner:2018zrj}. We indicated the favored and excluded regions for future $a_\mu$ measurements as the hashed green and gray regions.

\subsection{Cooling of SN 1987A}
\label{sec:sn}
A core-collapsed supernova (SN) behaves like a proto-neutron star and cools mainly through neutrino diffusion under the standard SN model. The measured SN 1987A neutrino burst flux agrees with SN model predictions~\cite{raffelt1996stars} and hence can be used to constrain light dark scalars, which may provide an extra cooling channel. For the extra cooling to be effective, dark scalars should be abundantly produced in the supernova and not decay or trapped by nucleons and leptons along their exiting path.  Reference~\cite{Dolan:2017osp} has considered an axion-like particle (ALP) with photon coupling $-(g_{\gamma \gamma}/4)a F \tilde{F}$.  The dominant production and trapping mechanisms are through a Primakoff process $\gamma N \to N S$ and an inverse-Primakoff process $S N \to N \gamma$, respectively. Given the production and trapping mechanisms for ALPs are identical to those of muon-specific dark scalars, we ignored the difference in the CP quantum number and directly translated the resulting SN bounds on $g_{\gamma\gamma}$ from~\cite{Dolan:2017osp} to those on $g_\mu$ for muon-specific dark scalar through the relation
\begin{equation}
g_{\gamma\gamma} = \frac{\alpha}{2 \pi} \left| \f{g_\mu}{m_\mu}F_{1/2} \left(\f{4 m_\mu^2}{m_S^2}\right) \right|.
\label{eq:ggammagamma}
\end{equation}
The translated bound is shown as the gray shaded region with label ``SN 1987A" in the right panel of Fig.~\ref{fig-result}~\footnote{The SN bound for the muon-specific dark scalar here is a factor of 2 lower than those in~\ref{fig-result}. We assumed $g_{\gamma\gamma}$ for the scalar and the pseudo-scalar to be identical while~\ref{fig-result} assumed the two couplings to differ by a factor of 2.}. Note that the exclusion region is cut off at $m_S= 2m_\mu$: a heavier scalar would be always trapped in the supernova core, without contributing to the supernova cooling~\cite{Batell:2017kty}.
 A similar analysis can be applied to the lepton-specific dark scalars except that the electron-dark scalar interaction should be also included for the trapping. Nevertheless, the resulting bound is weaker than the constraints of other electron beam-dump experiments for the parameter space of interest. Hence it is not explicitly labeled in the joint exclusion region in the left panel of Fig.~\ref{fig-result}. 

\subsection{Other fixed-target and beam dump experiments}\label{sec:complementary-dump}

In case of the lepton-specific model, a constraint to $g_\mu$ can be derived considering the emission of scalar particles by $e^+$ and $e^-$ in the beam-induced electromagnetic shower. Even if the corresponding radiative cross section is suppressed by the factor $(m_e/m_\mu)^2 \sim 2\cdot 10^{-5}$ with respect to the muon case, the much larger track-length of $e^+$ and $e^-$ compensates for this, resulting in a non-negligible yield.
In this case, since that the dump radiation length is much smaller than the target-detector distance, it is possible to assume that the scalar production happens always at the beginning of the target. This allows to exclusion limit for scalars from the sensitivity to a visibly-decaying dark photon~\cite{HOLDOM1986196,PhysRevD.86.095019}, by following the procedure depicted in~\cite{Chen:2017awl}. In the left plot of Fig.~\ref{fig-result}, results from E137 and BDX are reported.
 
In case of muon-specific model, there may be a significant contribution due to Primakoff process $\gamma N \to N S$ by bremmstrahlung photons in the beam dump. Although $g_\mu$ contribution to the $g_{\gamma\gamma}$ is loop-suppressed, a large flux of photons produced by the primary beam may still yield a large number of signal events and consequently a competitive bound on $g_\mu$. In particular, in Ref.~\cite{Dolan:2017osp} the original E137 bound on the mass and coupling of ALPs was cast to a limit on $g_{\gamma\gamma}$. In the calculation, authors used the photons track-length distribution reported in the original E137 paper~\cite{PhysRevD.38.3375}, 
and evaluated the experimental acceptance only including the angular effects due to the opening angle between the ALP decay products. We note the following critical issues in this procedure. First, it ignores the contribution to the detector geometrical acceptance from the finite production angle of ALPs in the beam dump. Secondly, we found the photons track length distribution reported in Ref.~\cite{PhysRevD.38.3375} to be inconsistent, in particular at low energy, with the result obtained from a \texttt{FLUKA}-based calculation that we performed (we cross-checked our result with the analytic predictions from showering theory~\cite{RevModPhys.13.240,Clement1965}, finding an excellent agreement).
We still report the result obtained translating the aforementioned limit on $g_{\gamma\gamma}$ to a limit on $g_\mu$, following the procedure depicted in Sec.~\ref{sec:sn}. We show the result as the grey-dotted line with label ``E137(sec. $\gamma$)" on the right panel of Fig.~\ref{fig-result}, leaving a more detailed 
study for the future. Note that we explicitly checked the momentum-squared of the $t-$channel photon from MC simulations and found it to be close to the on-shell condition, thus allowing us to use Eq.~\ref{eq:ggammagamma}.    

\section{Conclusions}
Muons play a special role in the Standard Model of particles and elementary interactions. Some discrepancies between the observations and the SM predictions may be an indication of new physics. In particular, the coupling  of muon to a new scalar particle ($S$) could reconcile some of the tensions between data and model. In this paper we elaborated on the idea that an $\mathcal{O} (10)\, \rm{GeV}$ electron beam hitting the dump is a copious source of muons and electron beam-dump experiments have the ideal  set up to detect the subsequent $S \to e^+ e^-$ and $S \to \gamma \gamma $ decays. A combination of simulation codes (\texttt{FLUKA} and \texttt{Geant4}) has been used to generate, propagate, and project muon-produced scalars on E137 and BDX detectors.  The numerical simulation realistically takes into account  the different processes involved: muons production from the primary electron, resulting energy and angular spread, $S$ production, including production vertex distribution, $S$ propagation and decay products distribution within the detector volumes. Results obtained for the E137 setup extended the exclusion limits, consistently with the reported null results, to cover a broader area in $g_\mu$ vs. $m_S$ parameter space. An even larger exclusion zone was obtained considering the planned BDX experiment. In conclusion, we demonstrated that electron beam-dump experiments (past and future) provide an enhanced sensitivity to new physics that may be specifically coupled not only to electrons or photons but also to muons.

\section*{Acknowledgments}
We thank Clive Fields for his help in determining the geometry and materials of the E137 experimental setup. AC and YZ thank Oliver Mattelaer for his help in computing the cross-section for the scalar radiative emission with \texttt{MadGraph5\_aMC@NLO 2.6.3.2}. AC thanks Brian Batell and Rouven Essig for permission in re-using a figure from their E137 paper. YZ thanks Brian Batell, Stefania Gori, Ahmed Ismail, and Felix Kahlhoefer for helpful discussions. MB and YZ thank the KITP at UCSB for hospitality, where their research was supported by the National Science Foundation under Grant No. NSF PHY1748958. YZ also thanks the Aspen Center for Physics for hospitality, where his research is supported by NSF grant PHY-1607611. LM acknowledges support from Universit\`a degli Studi di Genova. YZ acknowledges support from U.S. Department of Energy under Grant No. desc0015845.

\appendix

\section{Numerical computation of total cross-section and differential distributions for the scalar radiative emission process}\label{app:calchep}

The \texttt{Geant4}-based code we developed to simulate radiative $S$ production by muons (see next Appendix) requires as input the total cross-section $\sigma_\mathrm{TOT}$ for the process $\mu + N \rightarrow \mu+N+S$, as a function of muon kinetic energy, and the kinematic distribution of produced $S$ particles, as a function of scalar kinetic energy $T_S$ and production angle $\theta_S$. We computed numerically these quantities through the \texttt{CalcHEP 3.7}~\cite{Belyaev:2012qa} package. We implemented the aforementioned process in \texttt{CalcHEP}, by adding $S$ as a new particle, and including the corresponding interaction Lagrangian with muons, $\mathcal{L} = - g_\ell S \bar{\ell}\ell$ (see Eq.~\ref{eq:scalarL}). The model was further modified by means of the \texttt{userff.c} function, in order to account for the combined atomic and nuclear form factor $G_2(t)$.
We adopted the form-factor expression discussed in~\cite{PhysRevD.80.075018}. For most nuclei, $G_2$ is written as an incoherent sum of an elastic term and an inelastic contribution
\begin{align}
G_2 =&{} G_2^{el} + G_2^{inel} \\
G_2^{el}(t)=&{} Z^2 \left(\frac{a^2 t}{1+a^2t}\right)^2\left(\frac{1}{1+t/d}\right)^2  \\
G_2^{inel}(t)=&{} Z \left(\frac{{a^\prime}^2 t}{1+{a^\prime}^2 t}\right)^2 \, \cdot W_2^{p}(t)\\
W_2^{p}(t) =&{}
\left( \frac{1+\tau(\mu_p^2-1)}{(1+\frac{t}{0.71 \mbox{GeV}^{2}})^4} \; \; ,
\right)^2 
\end{align}
where $-t$ is the momentum-transferred squared, $\tau=\frac{t}{4m_p^2}$, $Z$ and $A$ are the nuclei atomic number and mass, $a=113 Z^{-1/3} / m_e$, $d=0.164 \,\mbox{GeV}^2 A^{-2/3}$, $a^\prime = 773 Z^{-2/3} / m_e$, $\mu_p=2.79$, and $m_p, m_e$ are the proton and electron mass. For the specific case of hydrogen atom, we adopted the following form-factor parametrization~\cite{PhysRevD.8.3109,Tsai:1973py}
\begin{align}
G_2^{el}(t) = & \left(1-F(t)\right)^2 \cdot W_2^{p}(t)\\
G_2^{inel}(t) = & (1-|F(t)|^2 ) \cdot W_2^{p}(t)\\
F(t) = & \left(\frac{t}{4\alpha^2 m_e^2}+1\right)^{-2}
\end{align}

\begin{figure*}[t]
\includegraphics[width=.95\textwidth]{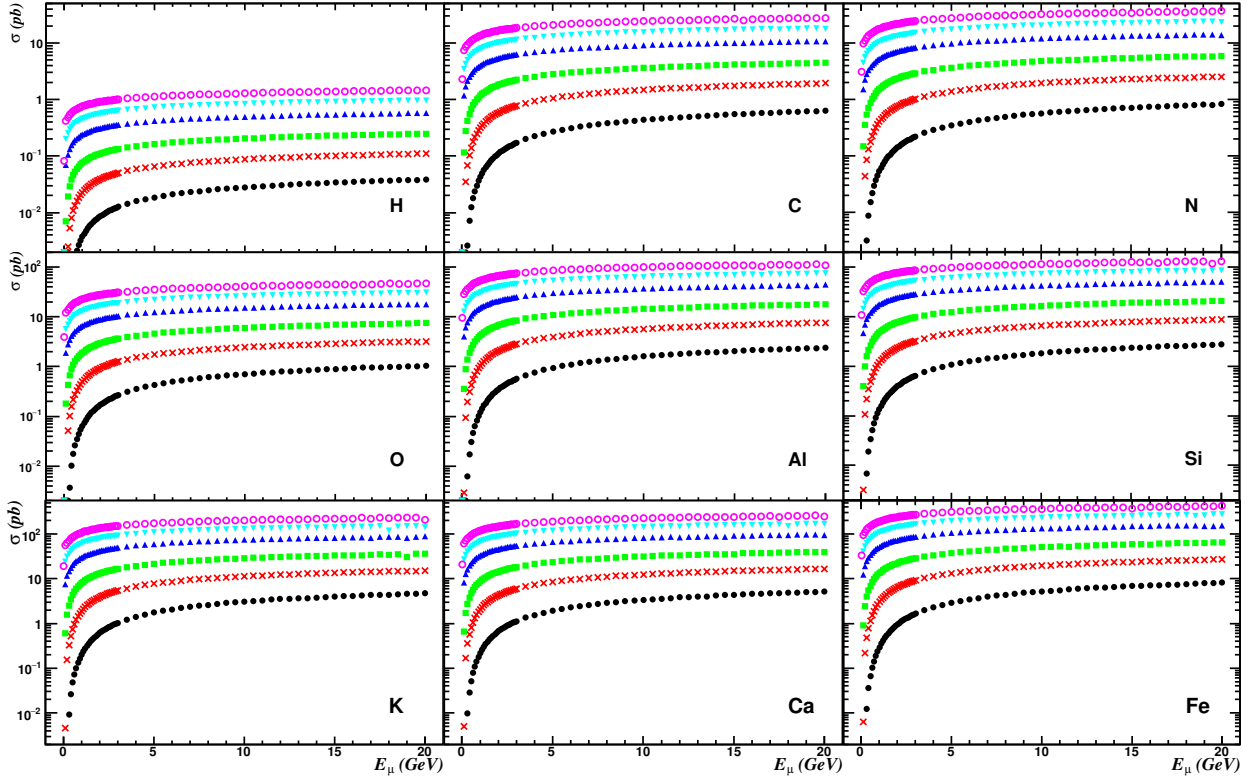}
\caption{Total cross section (in pbarn, y-axis) for the $\mu N \rightarrow \mu N S$ process as a function of the impinging muon kinetic energy (in GeV, x-axis), for $g_\mu = 3.87\cdot10^{-4}$. Each plot refers to a specific material, as reported in the corresponding caption. Different colors refer to different values of the scalar mass: 210 MeV (black full circles), 100 MeV (red crosses), 50 MeV (green squares), 20 MeV (blue triangles), 10 MeV (cyan triangles), 5 MeV (magenta open circles). Plots on the same line (column) share the same y-axis (x-axis) scale.
\label{fig:xsecdata}}
\end{figure*}

For each chemical element and $m_S$ value, we computed $\sigma_\mathrm{TOT}$ as a function of $E_\mu$ on a grid of 2000 points, from 10 MeV to 20 GeV. This interval corresponds to the energy range of interest in this work.
The value of the coupling was fixed to $g_\mu^0=3.87\cdot10^{-4}$. The numerical integration was performed by first adapting the \texttt{Vegas} grid with $it_1=10$ runs, each with $N_1=100$k iterations, then performing the actual calculation through $it_2=10$ runs, each with $N_2=100$k iterations. The target accuracy was set to $1\%$.
For each point, we also generated $100\cdot10^3$ Montecarlo events, that we used to sample the $T_S,\theta_S$ distribution. Figure~\ref{fig:xsecdata} shows the total cross-section for the $\mu + N \rightarrow \mu + N + S$ process, as a function of the impinging $\mu$ kinetic energy, for different values of the scalar mass, and for the materials that are more relevant in the simulation.
            
\begin{figure*}[t]
\includegraphics[width=.49\textwidth]{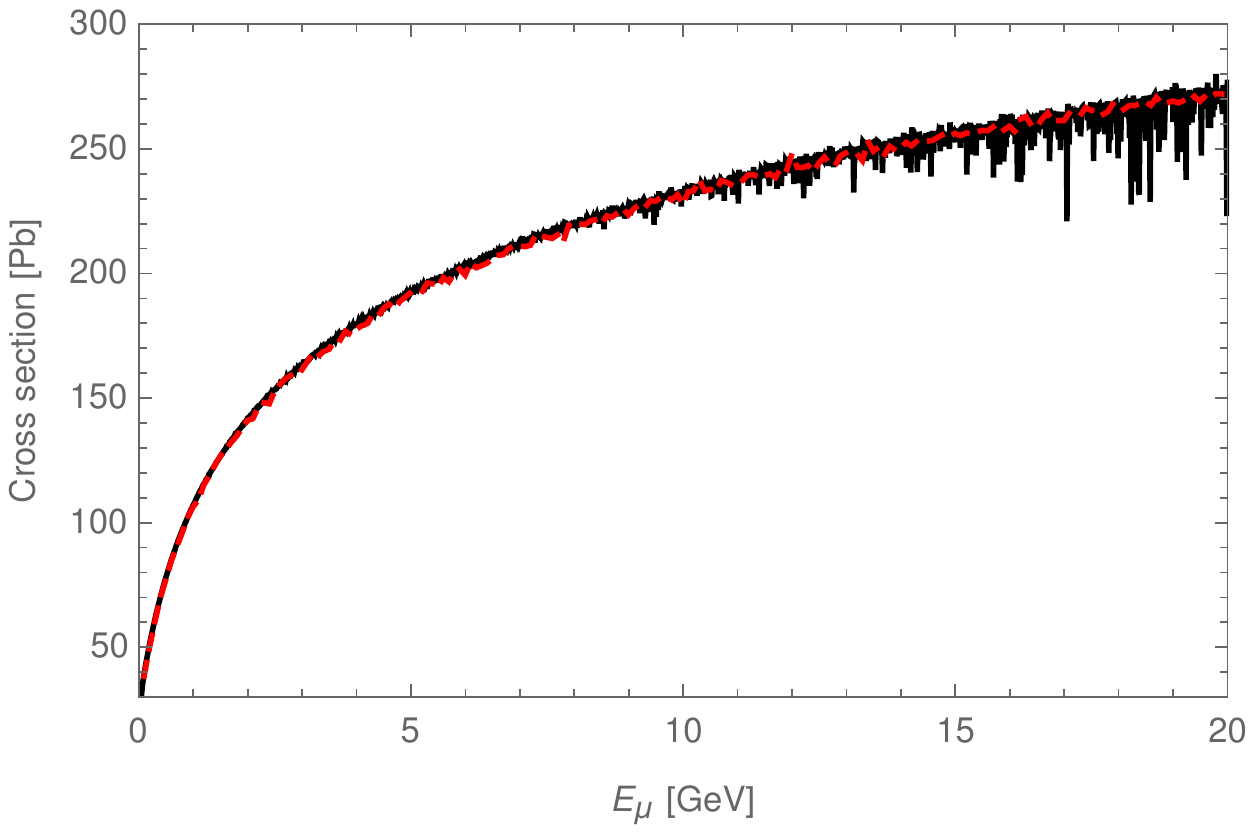}
\includegraphics[width=.485\textwidth]{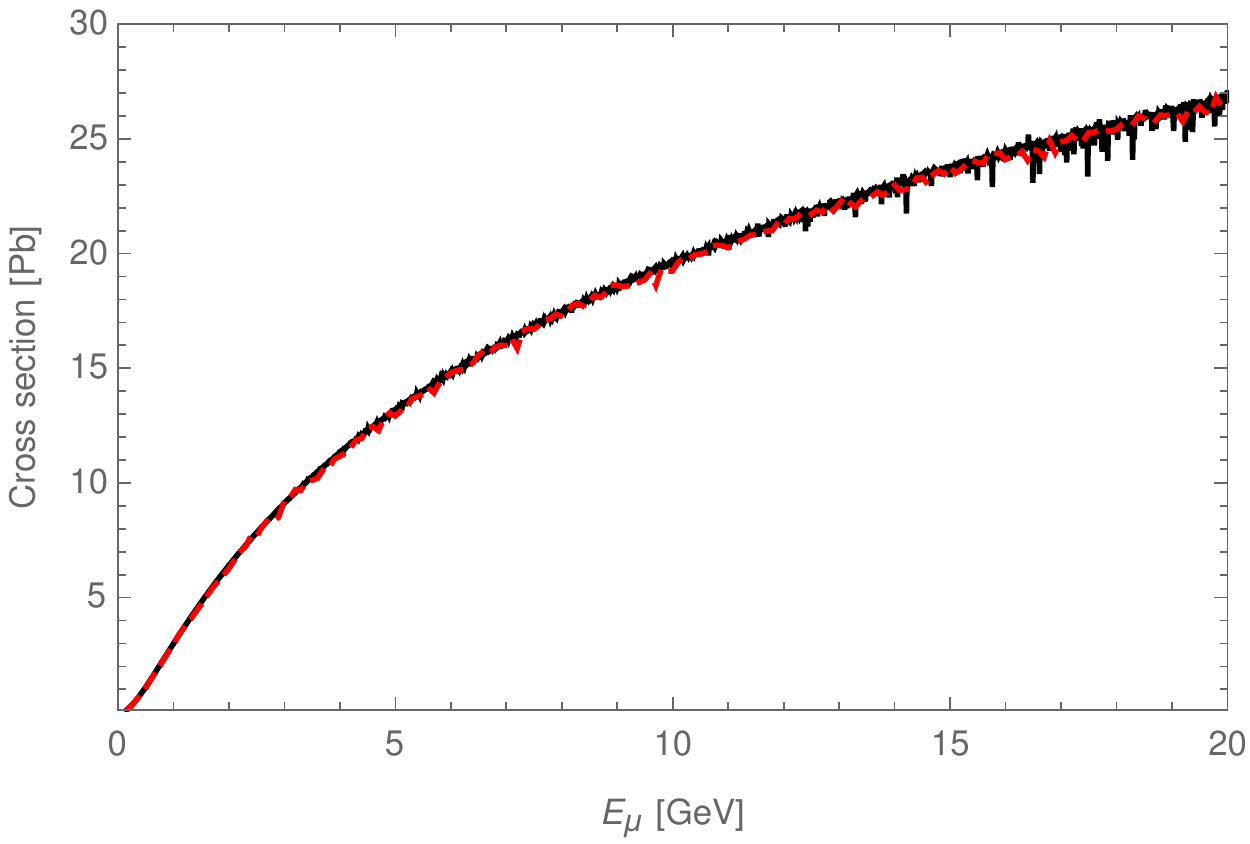}
\caption{Total cross section for the $\mu N \rightarrow \mu N S$ process for an iron target (for $g_\mu = 3.87\cdot10^{-4}$), as a function of the impinging muon kinetic energy, as obtained from \texttt{CalcHEP} (continous black line) and \texttt{MG5} (dashed red line). Left plot is for $m_S=10$ MeV, right plot for $m_S=100$ MeV.
\label{fig:comparison1}}
\end{figure*}

In order to confirm the numerical results, we compared them with those obtained from \texttt{MadGraph5\_aMC@NLO 2.6.3.2}~\cite{Alwall:2014hca} (\texttt{MG5} for short in below), finding an excellent agreement for all materials and scalar masses~\footnote{The computation of the cross-section for the scalar radiative emission process with \texttt{MadGraph5\_aMC@NLO 2.6.3.2} required an ad-hoc fix to the matrix-elements computation, to prevent a nonphysical discontinuity in the momentum transferred distribution.}.
As an example, Fig.~\ref{fig:comparison1} shows the comparison between the total cross section obtained from \texttt{CalcHEP} and \texttt{MG5} for Iron, as a function of the impinging muon kinetic energy, for the two values of the scalar mass $m_S=10$ MeV and $m_S=100$ MeV.

\section{\texttt{Geant} implementation of scalar radiative emission process}\label{app:g4}
  
We implemented the scalar radiative emission process by muons in \texttt{Geant4} as follows.
A new class \texttt{G4Scalar}, inheriting from \texttt{G4ParticleDefinition}, was introduced to describe the scalar particle $S$. The class contains a singleton \texttt{G4ParticleDefinition} instance, that effectively introduces the new particle in the simulation. The singleton is instantiated through a static method with a single parameter, the mass of the scalar particle. This allows to set $m_S$ dynamically at the simulation start, in order to perform the multi-step numerical evaluation of the event yield, as described in Sect.~\ref{sec:reach_procedure}.

The production process was implemented through the new class \texttt{G4MuonScalarProduction}, inheriting from \texttt{G4VDiscreteProcess} and implementing the following mother class virtual methods:
\begin{itemize}
\item \texttt{GetMeanFreePath}, that returns the mean free path associated to the $S$ production by a muon with given energy, in a given material.
\item \texttt{PostStepDoIt}, that specifies the actions to perform when a new scalar is produced, instructing the \texttt{Geant4} application to create a new scalar particle at the interaction vertex with given momentum, sampled from the differential cross-section, and to alter the impinging muon momentum to ensure total four-momentum conservation.
\end{itemize}

By making use of a class inheriting from \texttt{G4VDiscreteProcess}, and adding it to the list of allowed physics processes for muons in the main physics list, the \texttt{Geant4} kernel automatically handles the $S$ production coherently with the other muons interaction and propagation mechanisms.

The mean free path $\lambda$ for a muon with kinetic energy $T_\mu$ propagating in a certain chemical element is computed from the relation $\lambda^{-1} = {n\sigma_\mathrm{TOT}}$, with $n$ being the number of atomic nuclei per unit volume. The value of $n$ is obtained from the internal \texttt{Geant4} database, while $\sigma_\mathrm{TOT}$ is loaded by the \texttt{G4MuonScalarProduction} class at creation, from pre-computed data - more details are given in App.~\ref{app:calchep}. In case of compounds, $\lambda$ is obtained as an average:
\begin{equation}
\lambda^{-1} = \sum_{i} n_i \sigma^i_\mathrm{TOT}\; \; , 
\end{equation}
with the index ``$i$'' running over the chemical elements forming the compound, and $\sigma^i_\mathrm{TOT}$, $n_i$ being the corresponding cross-section and atomic number density .

To describe the emission of the new scalar particle, first the particle kinetic energy $T_S$ and polar angle $\theta_S$ with respect to the impinging muon direction are sampled from a pre-computed distribution, for the corresponding $E_\mu$ value and chemical element. The azimuthal angle $\phi_S$ is generated assuming a flat distribution. This allows to define the corresponding four-momentum $P_S$. The impinging muon four-momentum is thus altered to enforce four-momentum conservation, $P_\mu^\prime = P_\mu - P_S$ (this assumes that the target nucleus stays at rest during the interaction). In case of compounds, the chemical element through which the impinging muon interacts is randomly selected from those forming the element, associating to each a statistical weight $w_i = n_i \sigma^i_\mathrm{TOT}$. Then, the same procedure described before is adopted.


%

\end{document}